\documentclass[fleqn,12pt,twoside]{article}
\usepackage{espcrc1}
\usepackage{amsmath}
\usepackage{amssymb}

\usepackage{graphicx}
\usepackage[figuresright]{rotating}

\renewcommand{\AmS}{{\protect\the\textfont2
  A\kern-.1667em\lower.5ex\hbox{M}\kern-.125emS}}

\title{Kinetic equilibration in heavy ion collisions: 
the role of elastic processes}

\author{J. Serreau\address{Institut f\"ur Theoretische Physik der 
        Universit\"at Heidelberg,\\Philosophenweg 16, Heidelberg, Germany}
	\thanks{This talk is based on a work done in collaboration with 
	D.~Schiff~\cite{JSDS}}}
       
\begin{document}

% typeset front matter
\maketitle

\begin{abstract}
We discuss the question of thermalization during the very early stages 
of a high energy heavy ion collision. 
We review a recent study where we explicitely showed that, contrarily
to a widely used assumption, elastic collisions between the produced 
partons are not sufficient to rapidly drive the system toward local 
kinetic equilibrium. We then briefly discuss recent developments 
concerning the description of kinetic equilibration and comment on 
some open issues related to phenomenology.
\end{abstract}

\section{Introduction}

Present and upcoming high energy heavy ion collisions at RHIC and LHC
offer a great opportunity to test our understanding of the dynamics
of strong interactions in unusually extreme conditions.
A central question is that of the possible formation of a locally 
equilibrated state of deconfined matter, the so-called quark-gluon plasma 
(QGP). By now, there is no doubt that during the very early stages of such 
collisions the relevant degrees of freedom are quarks and gluons and it 
is rather clear that the initially produced partons undergo many subsequent 
rescatterings~\cite{SatzMueller}. It is a question of great interest 
for interpreting experimental data to know whether this dense gas of 
partons thermalizes before hadronization and, if yes, on which time scale. 

It is usually assumed that elastic scatterings between 
the produced partons, which randomize their momenta, rapidly drive the 
system toward a state of local kinetic equilibrium, on time scales 
$\lesssim 1$~fm~\cite{Biro,BMPR}. Under this assumption, the subsequent 
evolution toward local thermodynamic equilibrium consists in the equilibration
of the chemical composition of the gas through inelastic number changing 
processes~\cite{Biro}. However, the estimate of the kinetic equilibration
time scale on which the previous picture relies is not satisfying. First, 
it is based on the assumption of a gluonic system near equilibrium~\cite{BMPR}, 
whereas in an actual collision the produced gluons are far away from equilibrium. 
Second, it neglects the effect of longitudinal expansion at early times, 
which may have dramatic consequences: on the one hand, longitudinal 
free-streaming tends to destroy the isotropy of the momentum distribution;
on the other hand, expansion dilutes the system and the collisions 
between partons, which drive the system toward equilibrium, become more 
and more rare as time goes on. If the expansion is too strong, it may even 
be that the system never reaches equilibrium~\cite{HeisWang}.
Moreover, the importance of the initial condition which characterizes the 
partonic system just after the collision has been emphasized~\cite{HeisWang,Mueller}. Recent developments concerning a realistic description of the initial state 
have triggered more reliable studies of the microscopic evolution of the gluon
gas~\cite{Mueller,Raju,Gyulassy,Dumitru,JSDS,BMSS,Shin}. 
Here we report on a detailed study where we explicitely showed that the 
assumption of a rapid kinetic equilibration due to elastic collisions is 
not correct. We summarize the main hypothesis and results of this work; more 
details can be found in~\cite{JSDS}. We then discuss the question
of thermalization at the light of recent works.

\section{Kinetic equilibration: the role of elastic processes}

The matter produced in the central rapidity region is essentially made 
of gluons with transverse momentum $p_t \sim 1-2$~GeV~\cite{Mueller,minijet}. 
We compare the case where initial conditions are given by the saturation 
scenario, where the non-abelian Weizs\"{a}cker-Williams fields of the 
incoming nuclei materialize in on-shell gluons after the collision~\cite{Mueller}, 
to the case where the initial gluons are produced incoherently in semi-hard 
(perturbative) processes, the so-called minijet scenario~\cite{minijet}.

Following Mueller, we assume that already at early times a Boltzmann 
equation can be written for the partonic phase space distribution. Including 
only $2 \rightarrow 2$ elastic processes in the small-scattering angle limit, 
one obtains, in the leading logarithmic approximation, the QCD analog of the 
Landau collision integral~\cite{Mueller}. We use a relaxation time
approximation, which allows us to make further analytical progress.
In particular, we consider simple observables which directly probe
kinetic equilibration and for which all the phase-space integrals can be 
computed analytically. One is then left with a set of coupled one-dimensional
dynamical equations, which are easily solved numerically. We compute the time-dependent relaxation time in a self-consistent way and we carefully take 
care of conservation laws. Comparing with the results of~\cite{Raju}, 
where the exact solution is worked out numerically in the saturation scenario, 
one can assess the reliability of our approach, which appears to be quite good. 
We then apply this method to the minijet scenario~\cite{minijet,Dumitru}. 
We argue that, to characterize kinetic equilibration, a reliable criterium 
is to test the isotropy of various observables. As a consequence we find 
in particular that for both initial conditions the system does not reach 
equilibrium at RHIC energies. We test the robustness of our results 
by studying their sensitivity to the details of our description. We 
argue in particular that, due to the fragility of the weak coupling 
approximation, it appears difficult to obtain definite conclusions at 
LHC energies. In any case, we obtain reliable lower limits for the 
equilibration time. Requiring that the ratio of longitudinal 
and transverse pressures -- which should be $1$ in equilibrium -- be 
greater than $0.8$, we get:

\begin{tabular}{ll}
{\it - Saturation scenario:} & $t_{\rm elastic} \gtrsim 6$~fm at RHIC and 
$t_{\rm elastic} \gtrsim 3$~fm at LHC;\\
{\it - Minijet scenario:} & $t_{\rm elastic} \gtrsim 10$~fm at RHIC and 
$t_{\rm elastic} \gtrsim 5$~fm at LHC.
\end{tabular}

\noindent
For comparison, recall that, for example
at RHIC, the time during which a partonic description makes sense is 
at most of the order of $10$~fm.

It is interesting to quantify our results with respect to the analytic
estimate of the equilibration time obtained by 
Mueller in the saturation scenario~\cite{Mueller}: 
$t_{\rm elastic} \simeq c \, Q_S^{-1} \, \exp \sqrt{1/\alpha_S}$, where 
$\alpha_S$ is the strong coupling constant, $Q_S$ is the saturation scale, 
and $c$ is a numerical constant of order one. Using $Q_S$ as 
the scale fixing the value of the running coupling constant and inserting 
realistic values ($Q_S \simeq 1$~GeV at RHIC and $Q_S\simeq 2$~GeV at LHC) 
one gets $t_{\rm elastic} \simeq c$~fm at RHIC and $t_{\rm elastic} \simeq 
0.6 \, c$~fm at LHC. Comparing to our results, one concludes that $c \gtrsim 5$. 
This number, which is in principle of little importance in the weak coupling 
limit, plays a crucial role for values of the parameters appropriate for RHIC 
and LHC energies.

To conclude, we have shown that the kinetic equilibration time due to elastic
processes only is at least an order of magnitude bigger than the typical 
$1$~fm estimate usually assumed. Of course, this does not mean that 
kinetic equilibrium is not achieved in a real collision, but this 
implies that the standard picture of the space-time evolution of the system, 
described in the introduction, has to be modified. Important progress
are being made in this respect and a more reliable picture is already 
emerging. In the remaining sections, we try to summarize and discuss 
the current state of this rapidly evolving subject.

\section{Beyond elastic processes: the bottom-up scenario}

In the context of the saturation picture of gluon production, the 
authors of~\cite{BMSS} have pointed out the crucial role played by 
the dominant $2 \rightarrow 3$ branching process and have developed 
a complete description of kinetic equilibration: the so-called bottom-up 
scenario, which predicts that the equilibration time is parametrically 
given by: $t_{\rm eq} \sim Q_S^{-1} \, (1/\alpha_S)^{13/5}$. Inserting 
realistic values as before, on obtains typically: $t_{\rm eq} \sim 2$~fm 
both at RHIC and LHC. Moreover, from a comparison with the observed 
multiplicities at RHIC together with reasonable assumptions, the same 
authors infer the values: $t_{\rm eq} \simeq 3.2-3.6$~fm~\cite{BMSS}. 
Therefore, we see that, once $2 \rightarrow 3$ processes are taken into
account, kinetic equilibrium is likely to be achieved already at RHIC, 
however on a time scale which is still non negligible. This implies 
in particular that theoretical calculations addressing the various
experimental signatures of deconfined matter should take into account
the out of equilibrium evolution of the system~\cite{GSS}

To complete the picture, the important next step is to include
quark-antiquark pair production and to describe chemical equilibration. 
Previous studies indicate that the corresponding equilibration time is 
not small~\cite{Biro}, but this has still to be investigated in a bottom-up 
like picture. 

\section{Evidence for early thermalization in heavy ion collisions?}

Finally let us discuss a phenomenological puzzle, recently reported in the
literature~\cite{Heinz1}: experimental data on hadron spectra and elliptic 
flow appear to be very well described by perfect hydrodynamics, if one 
assumes that local thermal equilibrium is achieved already at a time 
$0.6$~fm after the collision.  As we have seen above, such a short time 
scale is however difficult to understand on the basis of perturbative QCD, 
which should be at least qualitatively reliable here.\footnote{The initial 
temperature required to match the data in~\cite{Heinz1} is $T\simeq 330$~MeV, 
which correspond to partons having a typical momentum $p\sim 3 T \simeq 1$~GeV.} 
Moreover, as pointed out by Mueller~\cite{SatzMueller}, this is at the
limit of applicability of an hydrodynamic description, this time being 
only slightly bigger than the time it takes for a typical gluon of this 
plasma to be produced. Nevertheless, the impressive agreement with 
data reported in~\cite{Heinz1} is striking and needs to be understood. 

One can follow two different directions:
the first one is to try to understand
how such a fast thermalization could occur. A recent suggestion
is that of non-perturbative processes leading to very rapid production 
of large amount of entropy~\cite{Shuryak}. Also, non-equilibrium 
phenomena may lead to so-called prethermalization, where some degrees 
of freedom rapidly exhibit a thermal behavior while the 
system as a whole is still out-of-equilibrium~\cite{pretherm,param}. 
Finally, another interesting idea is that the partons may already 
be ``thermalized'' in the incoming nuclear wave-function~\cite{KRZ}. 
The second direction is to investigate to which extent this agreement 
relies on the assumption of a completely equilibrated system. An
interesting study of this type has been made in~\cite{Heinz2}, where 
the authors considered the opposite limit, namely that of a system which 
is only transversally thermalized and which undergoes free-streaming in the 
longitudinal direction. Their results deviate from the ones obtained
with perfect hydrodynamics. However, it is not clear how fast the results
corresponding to intermediate situations would go from one limit to 
the other. In this respect, it would be very interesting to consider 
more realistic situations, where the system develops more and more 
longitudinal pressure as time goes on (see {\it e.g.}~\cite{JSDS}).

\end{document}